\documentclass[%
 reprint,
  amsmath,
  amssymb,
 aps,
]
{revtex4-1}

\usepackage{caption}

\usepackage{graphicx}
\usepackage{dcolumn}
\usepackage{bm}
\usepackage{booktabs}
\usepackage{multirow}
\usepackage{siunitx}

\usepackage{subcaption}
\usepackage{breqn}
\usepackage[table]{xcolor}
\usepackage[colorinlistoftodos]{todonotes}
\usepackage{lipsum}
\usepackage{adjustbox}

\begin{document}

\preprint{APS/123-QED}

\title{Multiple Fourier Component Analysis of X-ray Second Harmonic Generation in Diamond}
\author{P.Chakraborti$^1,$ B.Senfftleben$^2,$ B.Kettle$^3,$ S.W.Teitelbaum$^{4},$ P.H.Bucksbaum$^4,$ S.Ghimire$^{4},$ J.B.Hastings$^4,$ H.Liu$^4,$ S.Nelson$^{5},$ T.Sato$^{5},$ S. Shwartz$^6,$ Y.Sun$^4,$ C.Weninger$^{5},$ D.Zhu$^{5},$ D.A.Reis$^{4},$ M.Fuchs$^1$}
\affiliation{$^1$Department of Physics and Astronomy $,$ University of Nebraska-Lincoln$,$ Lincoln$,$ Nebraska$,$ USA}
\affiliation{$^2$Max Born Institute for Nonlinear Optics and Short Pulse Spectroscopy$,$ Berlin$,$ Germany}
\affiliation{$^3$Department of Physics$,$ Imperial College$,$ London$,$ UK}
\affiliation{$^4$PULSE Institute$,$ Stanford University$,$ Palo Alto$,$ California$,$ USA}
\affiliation{$^5$LCLS$,$ Stanford National Accelerator Lab$,$ Menlo Park$,$ California$,$ USA}
\affiliation{$^6$Department of Physics$,$ Bar$-$llan University$,$ Ramat$-$Gan$,$ Israel}

\begin{abstract}
The unprecedented brilliance of X-ray free-electron lasers (XFELs) \cite{Emma:2004a,Ishikawa:2012a} has enabled first studies of nonlinear interactions in the hard X-ray range. In particular, X-ray - optical mixing \cite{Glover}, X-ray second harmonic generation (XSHG) \cite{Shwartz.112} and nonlinear Compton scattering (NLCS) \cite{Fuchs} have been recently observed for the first time using XFELs. The former two experiments as well as X-ray parametric downconversion (XPDC)\cite{Freund:1969a, Eisenberger:1971a} are well explained by nonlinearities in the impulse approximation\cite{Eisenberger:1970a}, where  electrons in a solid target are assumed to be quasi free for X-ray interactions far from atomic resonances. However, the energy of the photons generated in NLCS at intensities reaching up to $4 \times 10^{20} \,\, \mathrm{W/cm^2}$ exhibit an anomalous red-shift that is in violation with the free-electron model. Here we investigate the underlying physics of X-ray nonlinear interactions at intensities on order of $10^{16} \,\, \mathrm{W/cm^2}$. Specifically, we perform a systematic study of XSHG in diamond. While one phase-matching geometry has been measured in Shwartz \textit{et al.}\cite{Shwartz.112}, we extend these studies to multiple Fourier components and with significantly higher statistics, which allows us to determine the second order nonlinear structure factor. We measure the efficiency, angular dependence, and contributions from different source terms of the process. We find good agreement of our measurements with the quasi-free electron model.

\end{abstract}

\pacs{Valid PACS appear here}
\maketitle


\section{\label{sec:level1}Introduction}

At optical wavelengths, second harmonic generation is well understood since the experiment by Franken et \textit{al.}\cite{Franken} in the 1960s. Classically, it can be explained by the nonlinear polarization induced by a strong oscillating electric field in an anharmonic atomic potential. In the dipole approximation it can be shown that second order interactions can only occur in non-centrosymmetric materials \cite{Boyd}. However, in the hard X-ray range, a different nonlinearity is dominant because of the unfavorable frequency scaling of the atomic nonlinearity, which for the second order polarization scales as $\omega^{-6}$ far from resonances. At X-ray frequencies well above atomic resonances, the electrons in a solid can be treated as quasi free in the impulse approximation \cite{Eisenberger:1970a}. In this case the nonlinear effects are based on the anharmonic motion of the quasi-free electrons under the influence of a strong electromagnetic wave, similar to a collisionless cold plasma \cite{Freund:1969a,Eisenberger:1971b}. Here, the second order polarization scales as $\sim \omega^{-3}$. For a sufficiently large electric field, the electron can gain a relativistic momentum within half a light cycle. The Lorentz force due to the magnetic field becomes non-negligible and can be treated as a perturbation\cite{Shen}. Under such conditions, the electron performs a 'figure 8' motion \cite{Vachaspati}, which in addition to an inhomogeneous charge density leads to the emission of radiation at harmonics of the incoming frequency. This nonlinearity is fundamentally different from that routinely used at optical wavelengths. For example the generation of second harmonic radiation is permitted even in centrosymmetric crystals. More importantly, it combines the potential of nonlinear optics with the atomic spatial resolution of X-ray wavelengths. Using nonlinear X-ray interactions, first experiments have been able to measure quantities otherwise not accessible, such as the induced local optical response with atomic resolution \cite{Tamasaku:2011a}. In addition to potential applications, the investigation of this nonlinear response is of fundamental interest. 
Here we investigate X-ray second harmonic generation, the prototypical second order nonlinearity. We perform a systematic study of the nonlinearity by observing several Fourier components of the second order response and different boundary conditions, such as Laue and Bragg geometries. This allows us to measure the second order structure factor in diamond. We find that it is consistent with the quasi-free electron model, that has been successful in describing X-ray - matter nonlinearities such as parametric down conversion(PDC) \cite{Eisenberger:1971a}, X-ray optical sum frequency generation(XOSFG) \cite{Glover} and non-resonant X-ray second harmonic generation experiment (XSHG) \cite{Shwartz.112}.  
In the model we treat the Lorentz force due to the magnetic field as perturbation. This perturbation scales with the normalized vector potential $a_0 = eE/(\omega mc)$, where $E$ is the Amplitude of the electric light field with frequency $\omega$, $e$ the electron charge, $m$ the electron mass, and $c$ the speed of light. Although for X-rays this factor is comparably small because of their high frequency (in our case $a_0 \approx 10^{-5}$), the plasma-like nonlinearity is significantly stronger than that due to the anharmonic atomic potential. We assume a plane wave of the form $\bold{E}({\bold{r},t})=\bold{e}\frac{E}{2}\exp[i(\bold{k}\cdot\bold{r}-\omega t)]$+c.c where $\bf{e}$ , $E$ and $\bf{k}$ are respectively the X-ray polarization vector, electric field amplitude, and wave vector of the incoming field. We expand the source current term $\bold{J}(\bold{r},t)$ in the free-electron approximation perturbatively, which at lowest (second) order leads to a current density oscillating at the second harmonic frequency ($2\omega$) given by \cite{Nazarkin.67}
 
\begin{dmath}
 \bold{J}_{2\omega}(\bold{r},t)=i\rho_{o}(\bold{r})\frac{e^{2}}{2m_{e}^{2}\omega^{3}} [(\bold{E\cdot}\nabla)\bold{E}+i\omega(\bold{E}\times\bold{B})]\\
   +i\frac{e^{2}}{m_{e}^{2}\omega^{3}}[\nabla\rho_{o}(\bold{r})\cdot\bold{E}]
   \bold{E}.
   \label{eqn:current}
\end{dmath}
 
 Here, $\rho_{o}(\bold{r})$ is the unperturbed electron charge density and $\bf{B}$ the magnetic field. The first term (displacement term) is trivially zero for SHG in a plane wave approximation. The second current term is due to the nonlinear Lorentz force and oscillates in the propagation direction along $\bf{k}$. The third term, known as the Doppler term, only contributes in a non-uniform plasma. In particular, a perfect crystal has a periodically modulated charge density, which can be expanded in a Fourier series in terms of reciprocal lattice vectors $\bold{G_{m}}$ as $\rho_{o}(\bold{r})=\sum\limits_{m}\rho_{m}\exp(i\bold{G_{m}}\cdot\bold{r})$. As described below, this allows us to phase-match the process, which significantly increases the efficiency. To calculate the generated SH signal, we consider only the contributions from the Lorentz and Doppler terms. Owing to the low efficiency of SHG we assume our pump is undepleted and the outgoing field of the second harmonic, $E_{SH}$ has a slowly varying envelope. Substituting the current term in the wave equation and making these additional assumptions, for a given reciprocal lattice vector $\bold{G_{m}}$ the wave equation describing SHG in a crystal simplifies to the equation below\cite{Yudovich:15}
\begin{dmath}
    \sin(\theta_{SH})\frac{\partial E_{SH}}{\partial x}+\cos(\theta_{SH})\frac{\partial E_{SH}}{\partial z}+ \frac{1}{v_{SH}}\frac{\partial E_{SH}}{\partial t}=\nonumber\\
    \sqrt{\frac{\mu_0}{\epsilon_0}}\frac{e^{2}\rho_{m}
    n_{2 \omega}E^{2}(\boldsymbol{r},t)}{8m_{e}^{2}\omega^{3}}{[\bold{k}+2(\bold{G_m \cdot e)e}] \\ 
    \cdot \bold{e_{SH}}} e^{i\Delta\bold{k \cdot r}}.
    \label{Weqn}
\end{dmath}
 
\begin{figure*}
 \centering
 \includegraphics[height=7.0cm,width=17cm]{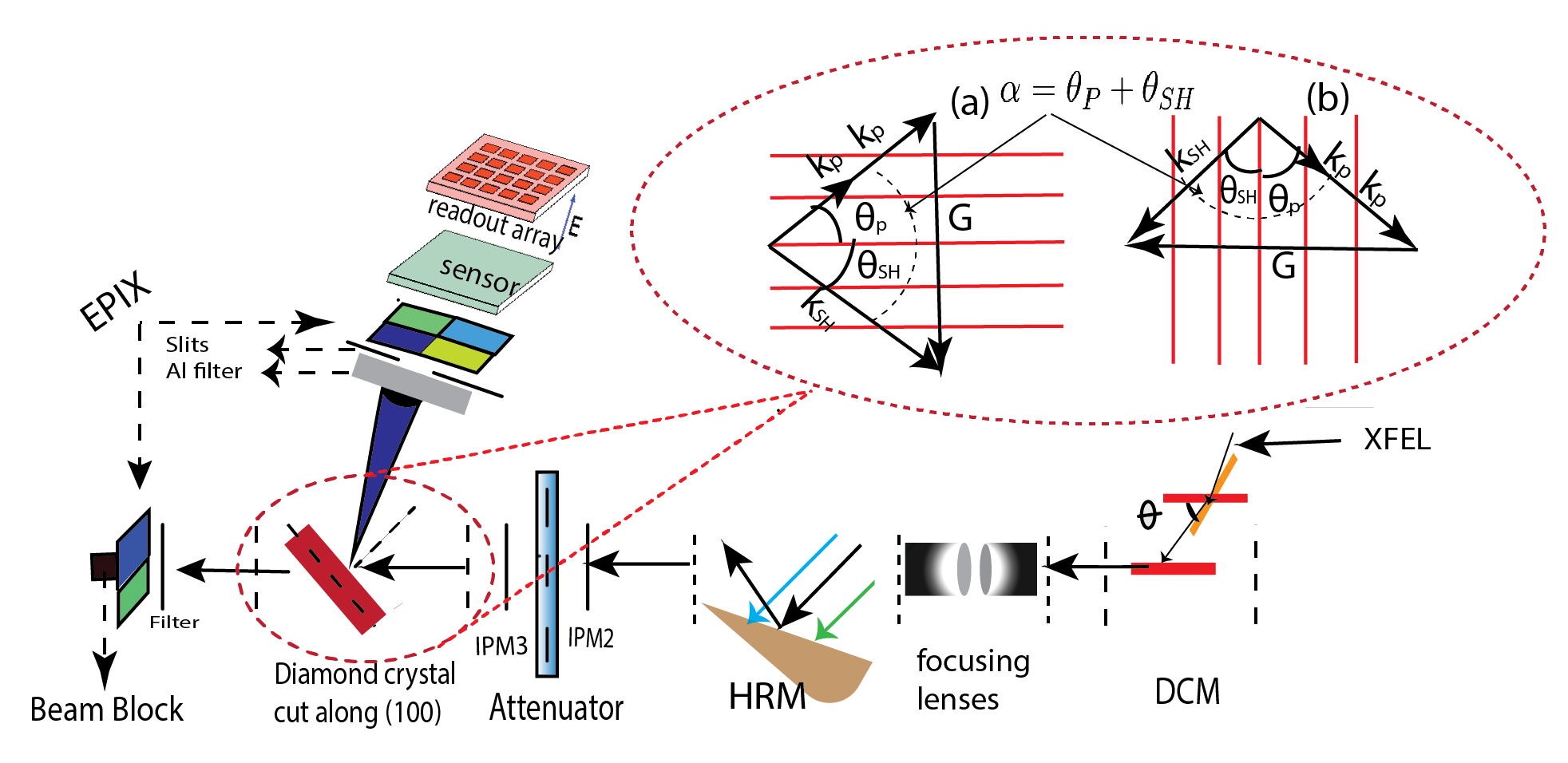}
 \caption{Experimental setup. The 9.8 keV XFEL beam is reduced to a bandwidth of $\sim$0.7 eV using a (111) diamond double crystal monochromator (DCM). The beam is focused to a waist size of 2.5$\mu$m using refractive lenses. Harmonic Rejection Mirrors (HRM) after the focusing lenses further suppress any residual second FEL harmonic radiation generated in the undulator by a factor of $10^{-4}$ to $10^{-5}$. The X-ray pulse energy on target can be varied using Si attenuators of different thicknesses. Intensity position monitors (IPM2 and IPM3) placed before and after the attenuators measure the total pulse energy of each shot and additionally monitor any drift in the beam position. The target used for second harmonic generation is a (100) cut diamond crystal. A 2-D EPIX pixel array camera installed on one arm of a diffractometer was used as a detector. It was possible to observe scattering angles covering $17^{\circ}$ to $124^{\circ}$, which allowed the investigation of multiple Fourier components of the SHG signal using both Bragg and Laue geometries for phase-matching. The inset shows the phase-matching condition for Bragg (a) and Laue (b) diffraction geometries, where $G$ is the reciprocal lattice vector of the planes shown, indicated in red, and $k_p$ and $k_{SH}$ are the fundamental and second harmonic photon wave vectors, respectively.}
 \label{fig:setup}
\end{figure*}

Here, $\theta_{SH}$ is the angle of the SH signal with respect to the chosen crystal lattice plane (see figure \ref{fig:setup}), $\epsilon_{o}$ and $\mu_{o}$ are electric and magnetic constants, respectively. $v_{SH}$ is the group velocity, $\bold{k_{SH}}$ the wave vector, $n_{SH}$ the refractive index, and $\bold{e}_{SH}$ the polarization vector of the generated second harmonic field.   We specify the $z$ coordinate along, and the $x$ coordinate transverse to the direction of the pump pulse propagation. $\Delta\bold{k}=2\bold{k}+\bold{G}-\bold{k_{SH}}$ describes the phase mismatch. The maximum signal can be achieved when the phase matching condition is satisfied, i.e. when $\Delta\bold{k} = 0$ (see figure \ref{fig:setup}). This condition leads to a small difference in angle from the linear Bragg diffraction of a $2\omega$ photon due to slight differences in twice the refractive index for the fundamental wavelength and the index of the generated second harmonic. The two terms in square brackets on the right hand side of the equation are the Lorentz and Doppler term, respectively. Depending on the geometry, which is given by the choice of the reciprocal lattice vector used for phase matching, the relative strengths of the two terms can vary as shown in Table IV below. As can be seen from equation \ref{Weqn}, the generated SH field scales quadratically with the field of the pump pulse, as expected for a second order nonlinear process.

The highly non-colinear geometry between the pump and the SH pulse limits the length over which the two beams overlap during the interaction. In particular, the small focus size and the ultra-short duration of the pump pulse leads to spatial and temporal walk-off between the pump and signal over the crystal length. Within a walk-off length the SH signal grows quadratically with the interaction distance. However, for our geometry, where the effective crystal thickness is much smaller than the Rayleigh length but significantly longer than the walk-off length, the signal grows overall only linearly with crystal length. For a monochromatic 1D X-ray field with a frequency far from resonances and which exactly fulfills the phase-matching condition ($\Delta \bold{k}=0$), the SH intensity emitted into the phase-matching direction for the $m$-th Fourier component is approximately given by  

\begin{dmath}
I_{2\omega} \approx \frac{c}{2 \epsilon_0 n_{2\omega}} \left[ \left( \frac{\bold{k}}{2} + \frac{(\bold{G_m} \cdot \bold{\hat e})(\bold{\hat e} }{k} \right) \cdot \bold{\hat e_{SH}}) \right]^2 a_{0,X}^4 \rho_{m}^2 L_{w_{X}} L_{\mathrm{c,eff}}.
\label{eqn:1D}
\end{dmath}
Here $a_{0,X}$ is the normalized vector potential for the pump pulse, $L_{w_{X}}$ the walk-off length, and $L_{\mathrm{c,eff}} = L/\cos\theta_p$ the effective crystal length, where $L$ is the crystal length.

For a quantitative comparison we numerically solve the wave equation (eq. \ref{Weqn}) for a temporally transform-limited Gaussian X-ray pulse with a finite bandwidth, and integrate over the length of the crystal. We then compare the theoretical efficiency of the SHG with the experimental value for different phase matching geometries.

\section{Experimental Setup}

The experiment was performed at the XPP instrument at the LCLS free-electron laser using an X-ray energy of 9.8 keV (see figure \ref{fig:setup}). The X-ray beam was monochromatized to a bandwidth of $\sim$0.7 eV using a double crystal diamond (111) monochromator for scattering in the polarization plane. The X-ray pulse with an energy of 178 $\pm$ 35 $\mu$J and a duration of 30 $\pm$ 10 fs (RMS) was focused to a spot size with waist radius of $w_0 = 2.5 \substack{+0.8 \\ -0.7} \, \mu$m, leading to an averaged intensity inside the focal spot of $I = 1.02 \substack{+11 \\ -2.8} \times 10^{16} \,\, \mathrm{W/cm^2}$. The pulse energy on target can be varied by introducing Si attenuators of different thicknesses into the unfocused beam. The second FEL harmonic emission generated in the undulator was significantly reduced before reaching the target by the structurally forbidden (222) reflection of the diamond monochromator, two reflections from gracing incidence hard X-ray offset mirrors and additional harmonic rejection mirrors. In our measurements essentially no linear diffraction from the second FEL harmonic signal from the target crystal was observed. As a target for second harmonic generation, we used a (100)-cut diamond crystal with dimension of 4 $\times$ 4 $\times$ 0.8 mm$^3$. We used various crystal lattice planes to investigate different geometries and multiple Fourier components of the charge density, both in Bragg reflection and in Laue transmission geometries. The crystal lattice planes and their respective scattering angles that were used as phase-matching geometries are listed in table \ref{Table:geometries}.

\begin{table}
\begin{center}
    \begin{tabular}{|l|l|l|l|l|l|} 
    \hline
      \textbf{Plane} & \textbf{Number of shots} & \textbf{$\theta_{G}$} & \textbf{$\theta_{P}$} & \textbf{$\theta_{\mathrm{Bragg}}$} & \textbf{$\theta_{SH}$}\\
      \hline
      (111) & 60k & 35.26 & 44.07 & 17.63 & 8.81\\
      \hline
      (11-1) & 126k & -35.26 & -26.44 & 17.64 & 8.82\\
      \hline
      (220) & 60k & 0 & 14.5 & 28.99 & 14.5 \\
      \hline
     (113) & 83k & 64.76 & 81.82 & 34.14 & 17.06\\
     \hline
     (00-4) & 122k & -90 & -69.27 & 41.47 & 20.73\\
     \hline
     (331) & 120k & 13.26 & 35.94 & 45.39 & 22.68\\
     \hline
     (660) & 77k & 0 & 48.67 & 97.4 & 48.67\\
     \hline
    \end{tabular}
    \caption{Geometries of the investigated SH Fourier components. All angle measurements are in degrees. $\theta_{G}$ is the angle between the respective reciprocal lattice vector used for phase-matching and the (110) lattice planes. $\theta_{P}$ is the angle of the incoming pump pulse and $\theta_{SH}$ is the angle of the outgoing SH signal, both measured with respect to the (110) plane as shown in Fig. \protect\ref{fig:setup}. $\theta_{\mathrm{Bragg}}$ is the Bragg/Laue angle for linear diffraction of a photon of energy $2\hbar \omega$. The incoming pump field is p-polarized with respect to the scattering plane of the crystal lattice.}
    \label{Table:geometries}
  \end{center}
\end{table}

As a detector, we used a pixelated 2D EPIX camera, which has an energy resolution of roughly 0.4 keV \cite{LCLS-HB}, which  allows us to distinguish between fundamental and second harmonic photons. However, the detector cannot distinguish between a single second harmonic photon with energy $2\hbar \omega$ and the pile-up of two or more lower-energy photons with a combined energy of $2\hbar \omega$ detected in the same pixel. Similarly, a pixel count registered at an energy of $4\hbar \omega$ could result from a single photon or from the pileup of two second harmonic photons detected in the same pixel. Due to the comparably small pixel size of 50 $\times$ 50 $\mu \mathrm{m}^2$, there is significant charge sharing between the pixels. We correct for this by implementing a 'droplet algorithm'\cite{Silke} on the observed images for each shot. To significantly decrease the background and the probability of pileup from the FEL fundamental, a 1.5mm thick Al filter was placed directly in front of the EPIX detector, which for the fundamental has a transmission of only $T_{\mathrm{9.8keV}} = 2.4 \times 10^{-5}$, while for the second harmonic $T_{\mathrm{19.6 keV}} = 0.3$.

The shot-to-shot pulse energy fluctuations of the XFEL beam are recorded using Intensity Position Monitor (IPM) diodes.These measurements allow us to more finely bin the intensity dependence of the measured SH signal as compared to only the Si attenuators. The on-target pulse energy was cross calibrated to the IPMs using an optical power meter.

\section{Experimental Results}
\begin{figure}
    \centering
    \includegraphics[height=7.2cm,width=9.1cm]{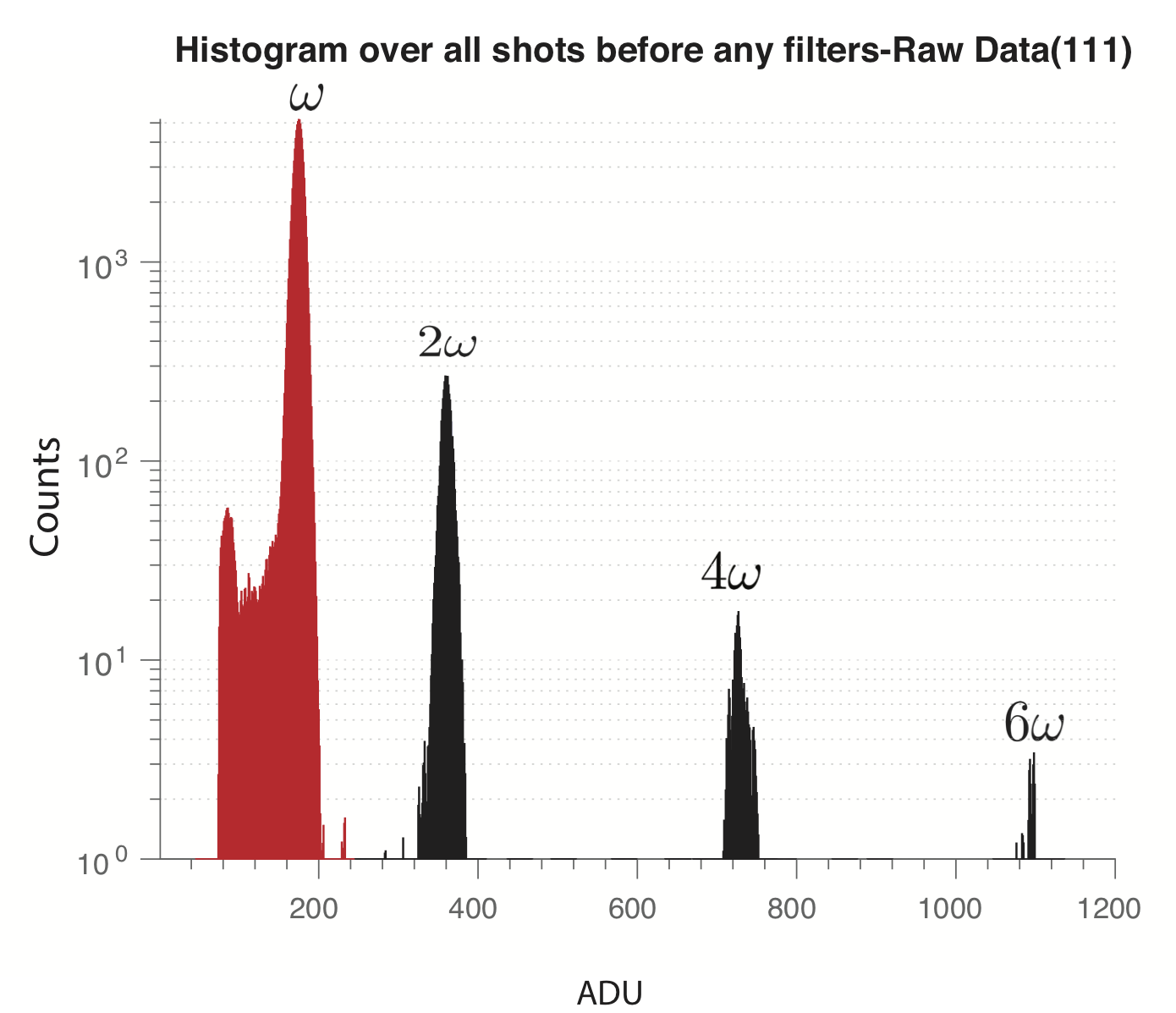}
    \caption{\textbf{Detector histogram of the observed signal from the (111) phase-matching geometry}. The histogram shows the number of detected photons as a function of photon energy, given in Analog-to-Digital-Units (ADUs), after running the droplet algorithm. A 9.8 keV photon corresponds to approximately 180 ADUs. The peaks corresponding to $4 \omega$ (720 ADUs) and $6 \omega$ (1080 ADUs) agree with pile-up statistics for second harmonic ($2 \omega$) photons. The histogram only includes the relevant small region of interest of the camera. To estimate the pileup from the fundamental photons, ($\omega$ $\sim$100 to 215 ADUs) within this small region of interest we  normalize the photon counts in the $2\omega$ bin ($\sim$200 to 430 ADUs) by the number of shots recorded for the geometry (TABLE I) and the number of pixels. The data shown here is not corrected for the detector quantum efficiency and the transmission through the Al filters in front of the camera.}
    \label{fig:histogram}
\end{figure}

The second harmonic signal is observed with a pixelated 2D array detector. Due to the comparably sparse signal and energy resolution of the detector, we can distinguish between photons with energies corresponding to the fundamental, second harmonic, and higher harmonics. A detector histogram for a run measuring the (111) Fourier component is shown in figure \ref{fig:histogram}. The SH peak at 360 Analog-to-Digital Units (ADUs) is sufficiently strong to generate pile-up signal at $4 \omega$ (720 ADUs) and $6 \omega$ (1080 ADUs). From these histograms we can estimate the pile-up contributions from both the fundamental and the second harmonic photons. These estimates are included when calculating negative and positive contributions to the error bars of the experimental data.

\begin{figure}
\centering
\begin{subfigure}[b]{0.5\textwidth}
	\includegraphics[height=6.5cm,width=7.5cm]{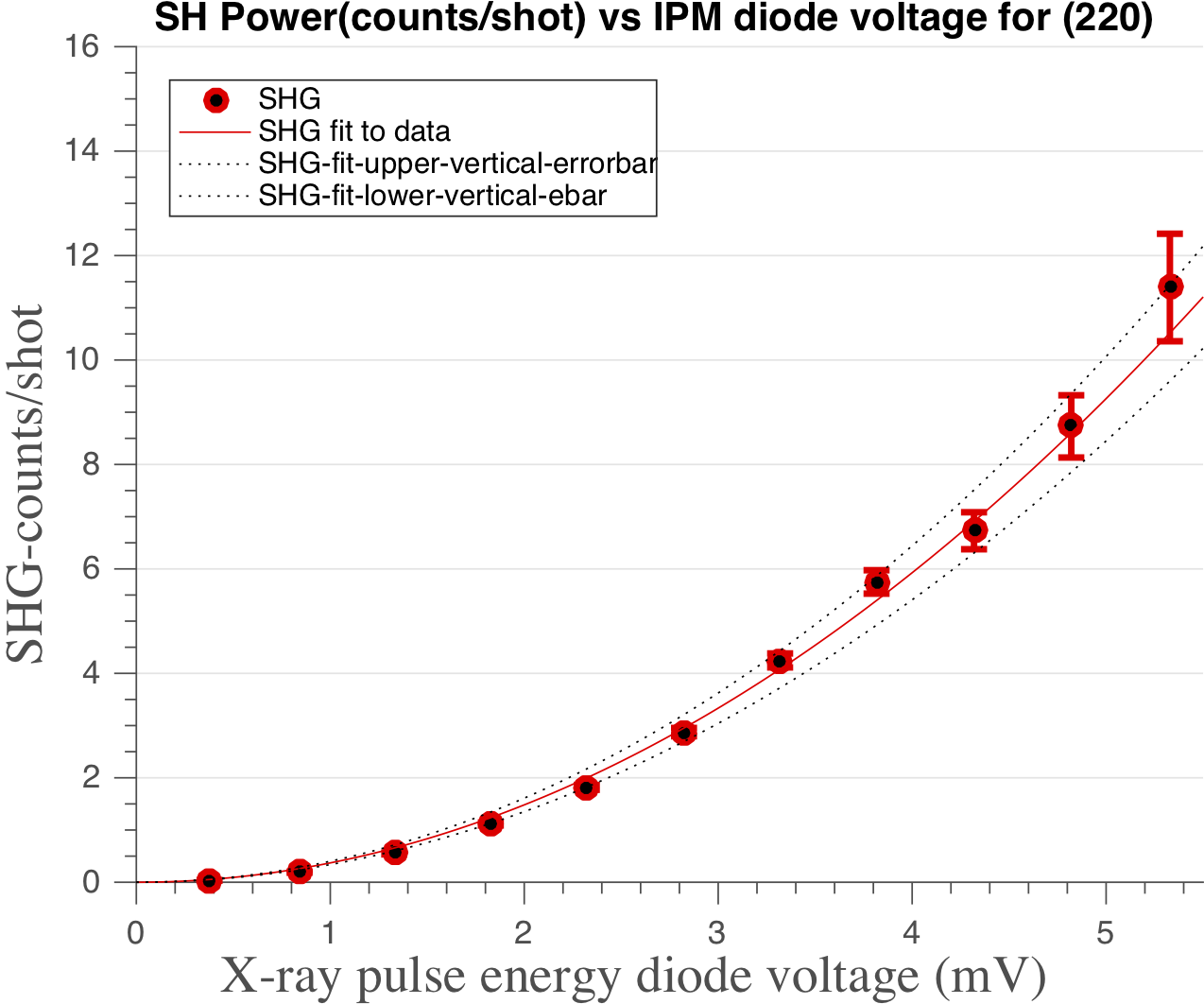}
\end{subfigure}
\begin{subfigure}[b]{0.5\textwidth}
	\includegraphics[height=6.5cm,width=7.5cm]{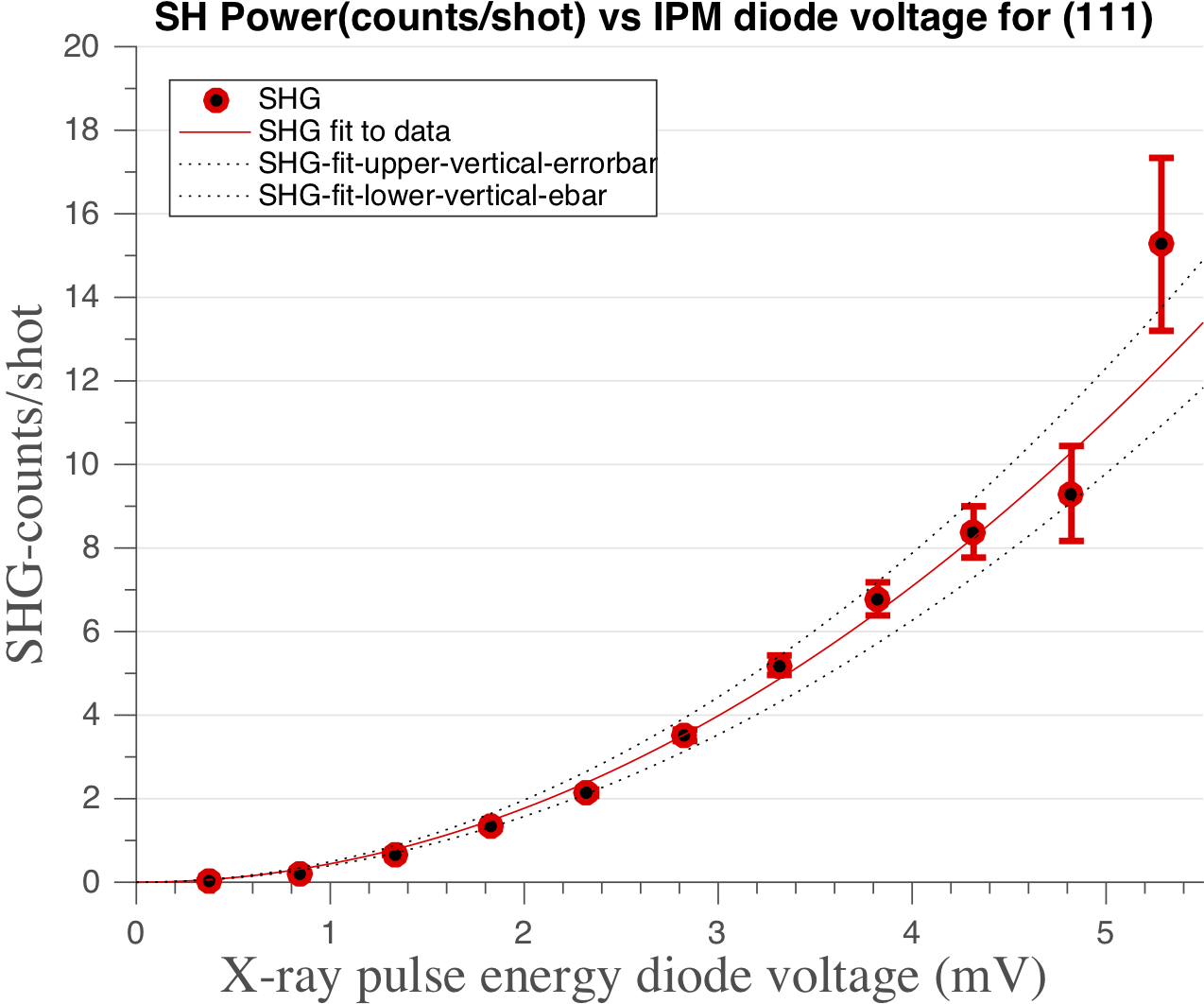}
\end{subfigure}
\caption{\textbf{X-ray pulse energy dependence of the measured SH signal}. The SH signal for the (220) and (111) reflections are plotted as a function of the X-ray pulse energy (IPM) diode voltage. They are normalized by the number of shots recorded within each pulse energy bin. The SH counts are corrected for the Al filter transmission and the camera quantum efficiency. The vertical error-bars are due to the counting error. The red curve shows a second order polynomial fit of the data and the dotted black curves are generated by fitting the upper and lower edges of the vertical error bars.}
\label{fig:Int-dependence}
\end{figure}

For each Fourier component we study the intensity-dependence and the dependence of the XSHG signal on small deviations from the phase-matching condition. The X-ray pulse energy is coarsely varied by using Si attenuators of different thicknesses. The FEL pulse energy on target varies from shot to shot due to the intrinsic energy fluctuations of the SASE lasing process. In the monochromator the shot-to-shot fluctuations in the FEL spectrum are also transformed into additional pulse-energy fluctuations. By measuring the FEL pulse energy after the monochromator for each shot using Intensity-Position Monitors (IPMs), we can use the fluctuations to achieve a finer pulse-energy binning. The dependence of the SH signal on the FEL pulse energy can be seen in figure \ref{fig:Int-dependence} for two representative phase-matching geometries. At the highest intensity, we observe comparably strong signals with several SH photons per shot after correction for the filter transmission and detector quantum efficiency. The number of generated photons show a quadratic dependence on the incoming X-ray pulse energy, as is expected for a second-order nonlinear process. To be more quantitative we fit the data with a second order polynomial function given by $n_{2\omega}=aV^{2}+bV+c$, where $n_{2\omega}$ is the number of second harmonic counts per pulse and $V$ is the IPM diode voltage in milli-Volt ($mV$). $a$ and $b$ have the units of $(mV)^{-2}$ and $(mV)^{-1}$ respectively. The first term in the equation is the usual quadratic dependence of a second order nonlinear process while the second and third terms take into account parasitic linear scattering of the FEL second harmonic and any background. Table \ref{Table:fitparameters} shows the resulting fit parameters for each measured Fourier component. We only list the coefficient for the quadratic dependence since the fit coefficients $b$ and $c$ were negligible for all geometries. The fitted SH signal shows a clear quadratic dependence on the incoming pulse energy for all geometries. At the highest intensities for some of the geometries $>$ 10 photons are generated per pulse, leading to efficiencies on order of $10^{-10}$. Since the SH signal is emitted into a narrow beam, these photons are concentrated into only a few detector pixels. This leads to a significant pile-up of SH photons, generating peaks at even higher orders in the detector ADU histograms (see figure \ref{fig:histogram}). Since the beamline transmission for photon energies higher than the FEL fundamental is extremely small and the higher-order peaks agree with Poisson statistics for pile-up, we also count these peaks as $2 \omega$ photons. We compare the measured efficiencies to theory in Table II.

\begin{figure}
\centering
\begin{subfigure}[b]{0.5\textwidth}
	\includegraphics[height=6.5cm,width=7.3cm]{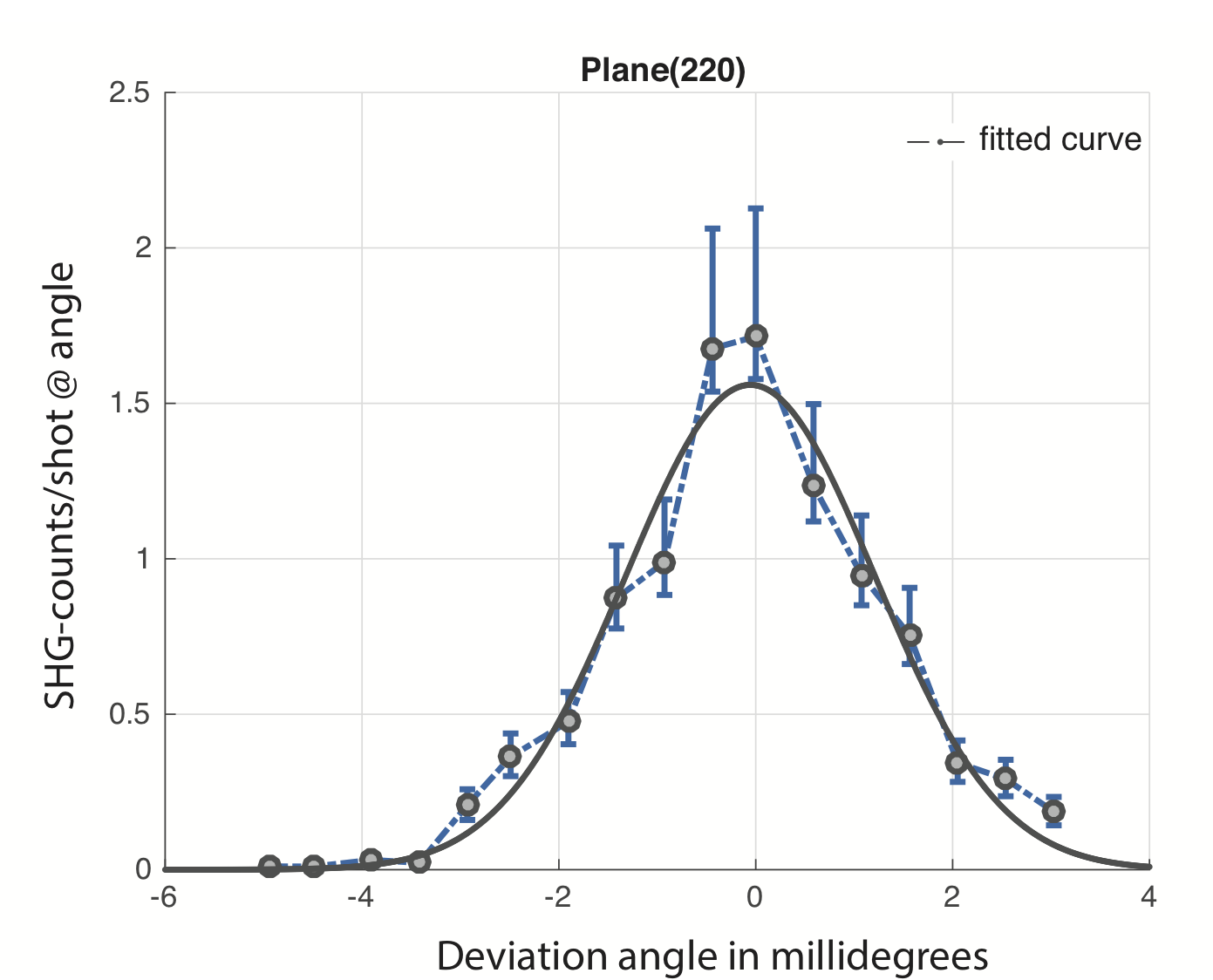}
\end{subfigure}
\begin{subfigure}[b]{0.5\textwidth}
	\includegraphics[height=6.5cm,width=7.3cm]{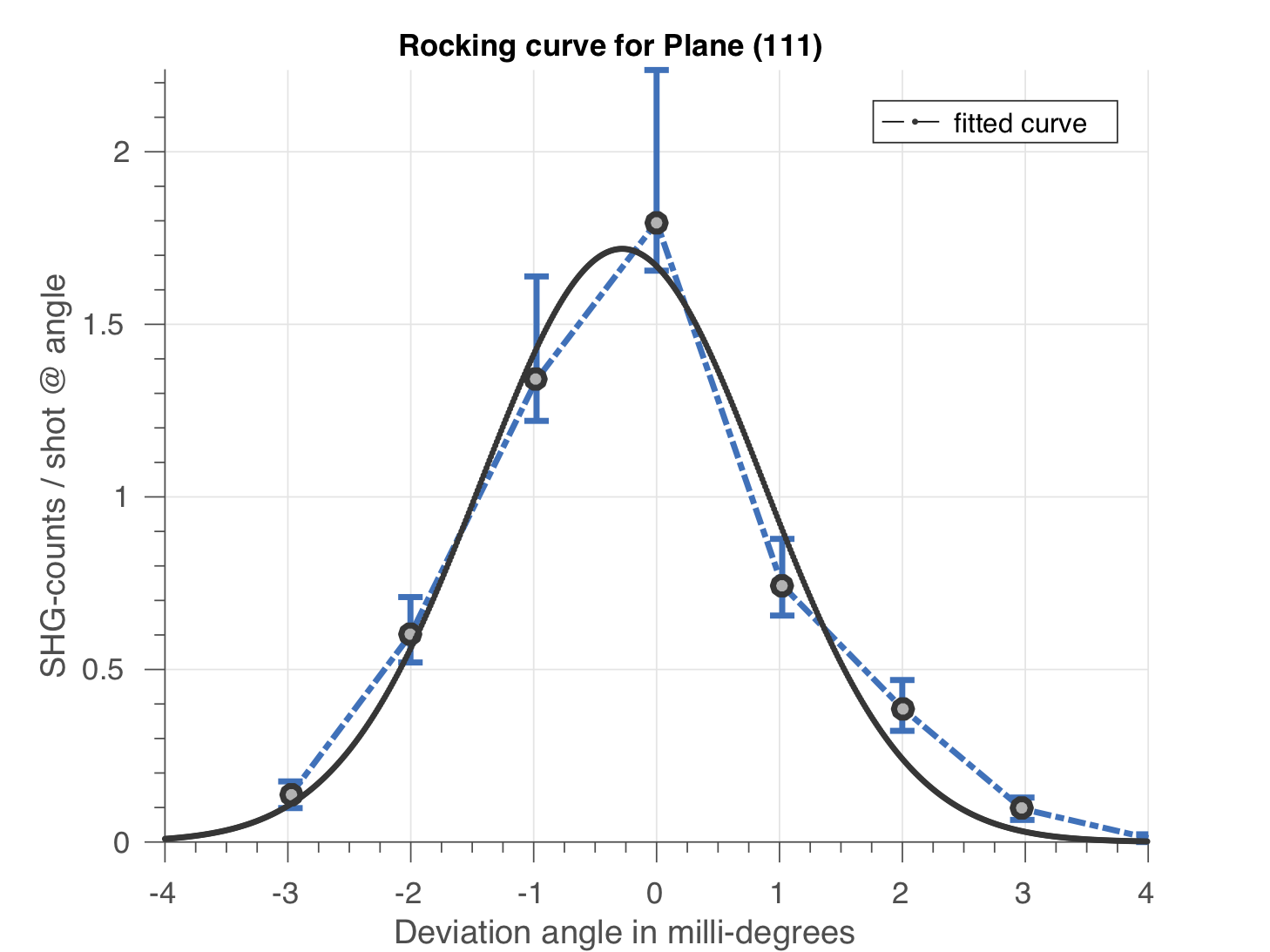}
\end{subfigure}
\caption{\textbf{Dependence of the second harmonic signal on the crystal rotation angle}. The generated SH signal is shown as a function of the angular deviation of the crystal angle $\theta$ from the phase matching angle $\theta_{SH}$. The curves only include shots within an IPM range of 1-2 mV in order to reduce the pile-up error. The solid black lines are Gaussian fits to the experimental data. \textbf{a)} shows the observed rocking curve for the (220) Fourier component and \textbf{b)} for the (111) component.}
    \label{fig:RC}
\end{figure}    

The generated SH signal as a function of the crystal rocking curve is shown in figure \ref{fig:RC} for the (220) and the (111) Fourier components. The observed signal has an angular offset from the expected linear elastic Bragg signal of the second FEL harmonic that agrees with the phase-matching condition described above. The SH signal is emitted into a narrow beam and is only well above background when the phase-matching condition is fulfilled. We calculate the width  for each Fourier component in table \ref{table:RC}). 
\begin{figure}
\centering
\begin{subfigure}[b]{0.5\textwidth}
	\includegraphics[height=7.3cm,width=9.0cm]{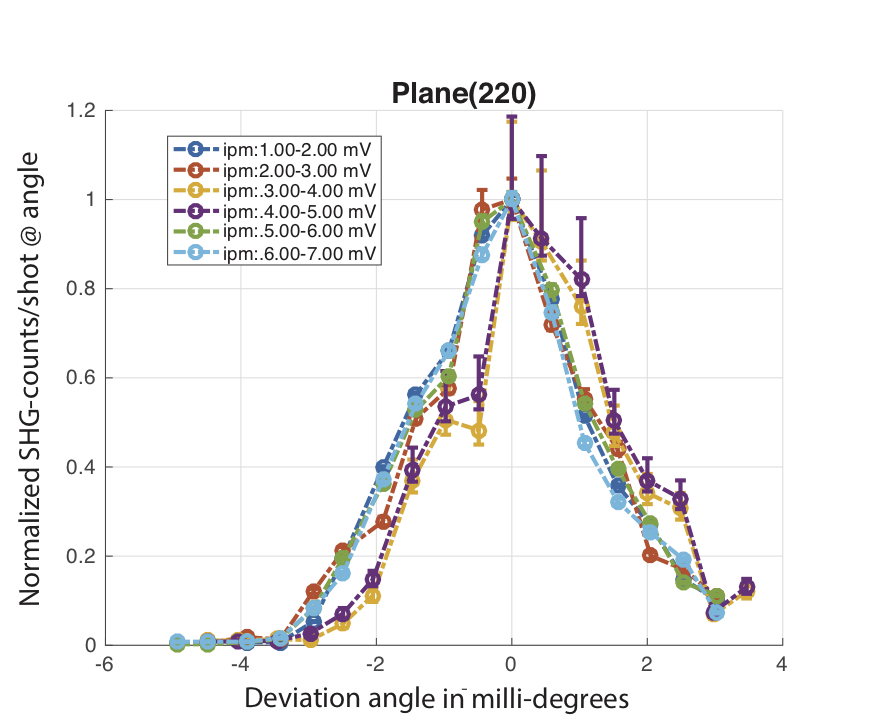}
\end{subfigure}
\begin{subfigure}[b]{0.5\textwidth}
	\includegraphics[height=7.3cm,width=9.3cm]{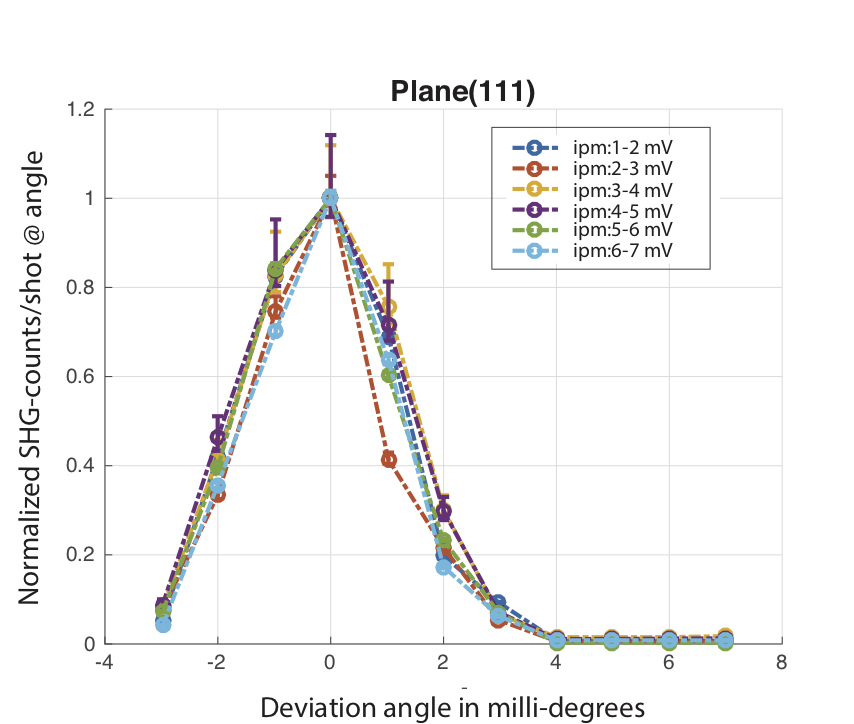}
\end{subfigure}
\caption{\textbf{Shape comparison of the crystal rocking curves for different ranges of X-ray Pulse energy} .The shape of the normalized curves for different FEL intensities are compared for the (220) (\textbf{a}) and the (111) (\textbf{b}) Fourier components. The FEL pulse energy is filtered by the intensity diode (IPM) values. We notice that the rocking curve widths are independent to any variations of the pulse energy}.
    \label{fig:Shape Comparison}
\end{figure}
 
\section{Discussion of results}

\begin{table}
 \caption{\label{Table:fitparameters} (a)Table showing the fit parameter $a$ from a polynomial fit for all eight XSHG Fourier components. We show here only the value of $a$, which is purely from the quadratic dependence of the incoming X-ray pulse, since the linear and constant fit parameters are negligible for all components. Also shown are the experimental and simulated efficiencies at the inelastic peak of the SHG signal. (b)Error-bars for the quadratic fit parameters, the simulated efficiency and the measured efficiencies respectively. }
 
    \centering
    \subcaption*{(a) Fitting Parameters and Comparison to Simulation}
    \begin{tabular}{|c|c|c|c|c|c}
   
    \hline\hline
    Plane  & $a$ & $\eta_{\mathrm{theory} }\times 10^{-10}$ & $\eta_{\mathrm{exp}} \times 10^{-10}$ \\
    
    \hline
    (111) & 0.5 & 7.3 & 1.42\\ 
    \hline
    (11-1) & 0.3 & 7.1 & 1.4\\
    \hline
    (220) & 0.4 & 6 & 1.1 \\ 
    \hline
    (113) & 0.07 & 3 & 0.66 \\
    \hline
    (00-4) & 0.2 & 2.7 & 0.52 \\
    \hline
    (331) & 0.05 & 3.2  & 0.61 \\
    \hline
    (660) & 0.02 & 0.5 & 0.11  \\
    \hline
    
    \end{tabular}
 \subcaption*{(b) Errors}
 \begin{adjustbox}{width=0.5\textwidth}
  \begin{tabular}{|c|c|c|c|c|c}
   
    \hline\hline
    Plane  & $a_{\mathrm{error}}$ & $\eta_{\mathrm{theory-error} }\times 10^{-10}$ & $\eta_{\mathrm{exp-error}} \times 10^{-10}$  \\
    \hline
    (111) &(+.05,-0.01) & (+2.1,-1.4) & (+1.5,-1.2) \\ 
    \hline
    (11-1) & (+0.05,-0.03) & (+2.1,-1.4) & (+1.52,-1.51) \\
    \hline
    (220) & (+0.03,-0.04) & (+2.,-1.3) & (+1.3,-1.1)  \\ 
    \hline
    (113) & (+0.02,-0.01) &(+1.6,-1.4) & ($\pm$0.3 ) \\
    \hline
    (00-4) &(+0.03,-0.02) &(2.1,-1.5) & ($\pm$0.5 )   \\ 
    \hline
    (331) & (+0.01,-0.01) &(+1.7,-1.5)  & ($\pm$0.31 )\\
    \hline
    (660) & (+0.02,-0.01) & ($\pm1.6$) & ($\pm$0.6 )  \\
    \hline
   
 \end{tabular}
 \end{adjustbox}
\end{table}
 
In order to explore the validity of the quasi-free electron model, we compare our measurements with theoretical results from the numerical solution of equation \ref{Weqn}. The different geometries also allow us to investigate the contributions of the different source terms described in equation \ref{eqn:current}(see TABLE III). We assume a temporally transform-limited Gaussian pulse and take into account the  pump field depletion during the process from the long attenuation length in diamond (1.2 mm for 9.8 keV photons). For each geometry we integrate the wave equation along the effective propagation length of the pump pulse inside the crystal. For the integration we use an interval significantly smaller than the respective spatio-temporal walk-off length. From the obtained outgoing SHG field, $E_{SH}$, we calculate the theoretical efficiency according to 
\begin{dmath}
 \eta_{\mathrm{SHG}}=I_{\mathrm{SH}}/I_{\mathrm{fund}},
\end{dmath}
where $I_{\mathrm{SH}}$ is the integrated second harmonic intensity at the exit of the crystal and $I_{\mathrm{fund}}$ the intensity of the incoming FEL fundamental. In table II(a) we compare the experimentally measured efficiencies to the numerical results for the highest FEL intensities. In case of the experimental values, the efficiency is given by the ratio of the number of observed SH photons corrected for filter transmission and detector quantum efficiency and the number of incoming FEL photons per pulse.  We find that the highest experimental efficiencies, as well as the simulated efficiencies for our phase-matching condition,$\Delta\bf{k}$ arise from the (111) and (11-1) planes in asymmetric Laue, and the (220) plane in symmetric Laue geometry(Table-II(a)). The efficiencies are of the order of $10^{-10}$ for both the simulation and the experiment. However, the theoretical results are approximately a factor of 5-6 higher than the experimental efficiencies.

A simple calculation of the efficiency from the 1D analytical plane-wave solution  (eqn. \ref{eqn:1D}) significantly overestimates the efficiencies($\sim 10^{-8}$), but it gives us a lot of qualitative insight into the SHG process. Due to the highly non-colinear geometry the SH signal gain is limited by walk-off. As the walk-off length $L_{w_{X}}$ is much smaller than both the Rayleigh range and the crystal thickness, the generated SH intensity scales bilinearly with $L_{w_{X}}$ and effective crystal length $L_{c,\mathrm{eff}}$ as $I_{\mathrm{SH}} \propto I_p^2 L_{w_{X}} L_{c,\mathrm{eff}}$. For our geometries we are mainly limited by spatial walk-off, given by $L_{w_{X}} = 2w_0/\tan(\alpha)$, where $2w_0$ is the focus diameter and $\alpha$ the angle between the pump and the generated SH signal. The efficiency scales with the FEL pulse properties as $\eta_{\mathrm{SHG}} \propto U/(\tau w_0)$, where $U$ is the pulse energy and $\tau$ the pulse duration. The relative efficiencies at similar intensities for the different Fourier components can be estimated by multiplying the source current terms with the walk-off lengths for each geometry. Specifically, in the plane wave case, the exactly phase-matched SH intensity is approximately given by $I_{SH} \approx 1/(2\epsilon_0 n_{2\omega}c) |J_{2\omega}|^2 L_{w_{X}} L_{c,\mathrm{eff}}$, which is essentially equation \ref{eqn:1D}. The magnitude of the two main source current density terms for each phase-matching configuration, namely the Lorentz and Doppler term in equation \ref{eqn:current}, are given in table \ref{table_current}. The listed current terms are the projections onto the SH polarization for each geometry. It can be seen that for most geometries the SH signal is mainly caused by the Doppler term. The high efficiencies of  the (220) and (111) components can be explained by the comparably strong current terms and long walk-off lengths. For the (660) Fourier component the SH emission angle is close to 90$^\circ$ to the pump pulse propagation, along the FEL polarization direction. Although linear scattering in this direction using p-polarized light is highly suppressed, a substantial SH signal can be observed. For this component the Lorentz current term oscillating along the FEL propagation direction is larger than the Doppler term. We do not observe any unexpected behavior of the SHG for the geometry where the signal is mainly due to the Lorentz term compared to the ones where the signal is mainly due to the Doppler term.

\begin{table}
\caption{Projected current density terms. The magnitudes of the Lorentz terms $J_{\mathrm{Lorentz}}$ and the Doppler terms $J_{\mathrm{Doppler}}$ from equation \ref{eqn:current} projected onto the polarization of the SH emission are listed in units of A/m$^2$. \label{table_current}}
    \begin{tabular}{|l||l|l|}
    \hline
    {\bfseries Plane} 
     & $J_{\mathrm{Lorentz}} \times 10^{-15}$ & $J_{\mathrm{Doppler}} \times 10^{-15}$ \\
     \hline
     (111) & 0.45 & 1.8\\
     \hline
     (11-1) & 0.45 & 1.72\\
     \hline
     (220) & 0.7 & 2.34 \\
     \hline
     (113) & 0.5 & 1.6\\
     \hline
     (00-4) &0.73 & 2.19\\
     \hline
     (331) & 0.6 & 1.4\\
     \hline
     (660) & 0.63 & 0.33\\
     \hline\hline
    \end{tabular}
\end{table}    

  The difference between experiment and simulation can be attributed to our assumptions of a monochromatic beam and perfect crystal quality. The error in the simulations is mainly due to uncertainties in the FEL parameters, specifically the FEL pulse energy, pulse duration, pulse structure, and focus size. The focus size and pulse duration not only affect the magnitude of the nonlinear source current term through the FEL intensity but also the walk-off lengths, over which the SH signal grows quadratically with distance. In order to account for this coupled error we have calculated the peak efficiencies for a set of FEL parameters within our uncertainties. From these numerical calculations we determine the upper and lower bounds for our error for each Fourier component. In particular we estimate the upper error bound using the highest X-ray pulse energy (216 $\mu$J), smallest focus size (1.8 $\mu$m), and shortest pulse duration (20 fs). The lower error bound is estimated at the lowest X-ray pulse energy (144 $\mu$J), largest focus size (3.3 $\mu$m), and longest pulse duration (40 fs). These results agree with the scalings obtained from the 1D model described above.

\begin{table}
\centering
\caption{Comparison of the FWHM-width and the integrated area (IA) in units of mdeg of the SH rocking curves for the measured Fourier components}
\label{table:RC}
\begin{tabular}{|l|{c}|S|S|}
    \toprule
    \hline

    \multirow{2}{*}{$\bf{G}_{m}$} &

      \multicolumn{2}{c|}{Experiment} \\
      & {IA} &  {FWHM}  \\
    \hline
      \midrule
      \hline
    (111) & 2.8  & 2.5  \\
    \hline
    (11-1) & 3 & 2.1 \\
    \hline
    (220) & 3  & 2.9  \\
    \hline
    (113) & 3.6  & 1.9  \\
    \hline
    (00-4) &4.2  & 4  \\
    \hline
    (331) & 3.2  & 3  \\
    \hline
   (660) & 3.3 & 4.2 \\
    \hline
    \bottomrule
  \end{tabular}
  \end{table}

We also investigate the dependence of the generated SH signal for small deviations from the phase-matching condition. To this end we measure the SH signal as a function of the crystal rotation angle. In order to examine the shape of the crystal rocking curves we filter the data by FEL pulse energy. The curves with the highest statistics (IPM range of $V_{\mathrm{IPM}} = 1-2$ mV) for the (111) and (220) Fourier component are shown in figure \ref{fig:RC}. The shape of the rocking curves do not show a significant dependence on the FEL fundamental intensities, as can be seen in figure \ref{fig:Shape Comparison}. Although the width of the observed rocking curves vary slightly for each Fourier component, the integrated reflectivities are nearly constant and within our uncertainties (see table \ref{table:RC}). The measured rocking curves are significantly broader than what is expected from the linear X-ray diffraction crystal Darwin widths. The broadening is mainly due to the divergence of the incoming converging beam, which is considerably larger than the crystal Darwin widths for any of the reflections. For a beam divergence of ($\sim 65 \, \mu \mathrm{rad}$)(4 mdeg) we expect rocking curve widths of $\sim 3.8 \mathrm{mdeg}$, which agrees reasonably well with the measured values and the integrated reflectivities.

\section{Conclusion}
We have experimentally investigated phase-matched X-ray second harmonic generation in diamond at a photon energy of 9.8 keV. We observe clear nonlinear signals with efficiencies on order of $\sim10^{-10}$ at X-ray intensities of  $I = 1 \times 10^{16} \,\, \mathrm{W/cm^2}$ for multiple Fourier components of the second harmonic response. For each component we measure a quadratic dependence of the signal on the incoming FEL intensity, as expected for a second-order nonlinear process. Observable signals are only generated in well-defined geometries around the phase-matching condition. We have compared the measurements to a theoretical model based on the impulse approximation, which assumes the electrons in a solid to be quasi free during the interaction with X-rays. In particular, we have compared the measured efficiencies, phase-matching condition, and slight angular deviations from perfect phase-matching. The calculated efficiencies are roughly a factor of 5 higher than the experimental values, which is in reasonably good agreement within the parameters assumed for the simulation. The relative intensities of the different Fourier components also agree with theory. The intensity of the SH signal rapidly drops once outside the phase-matching condition. The widths of the crystal rocking curves is mostly due to the X-ray beam divergence. From these comparisons we can conclude that our measured second order structure factor is consistent with the quasi-free electron model within the uncertainties of the experimental values and the approximations of our calculations. This experiment gives valuable insights into a nonlinear response that is fundamentally different from that routinely used at optical wavelengths. The elementary understanding of these processes is of great importance as these nonlinear processes in the X-ray wavelength range allow the investigation of non-equilibrium dynamics with atomic time and length scales. Despite low conversion efficiencies, these nonlinear interactions have the potential to lead to novel diagnostics with capabilities beyond linear diffraction techniques, especially considering the availability of current and future high-repetition rate FEL facilities.

\section{Acknowledgements}
This material is based upon work supported by the U.S. Department of Energy, Office of Science, Basic Energy Sciences under Award Number DE-SC0016494 and by the AMOS program within the Chemical Sciences, Geosciences, and Biosciences Division of Basic Energy Sciences, U.S. Department of Energy under Award Number DE-AC02-76SF00515. Use of the Linac Coherent Light Source (LCLS), SLAC National Accelerator Laboratory, is supported by the U.S. Department of Energy, Office of Science, Office of Basic Energy Sciences under Contract No. DE-AC02-76SF00515.

\bibliography{main}

\end{document}